\documentclass{aastex6}
\usepackage{mkfig}

\begin{document}

\title{{\em Hubble Space Telescope} Constraints on the Winds
  and Astrospheres of Red Giant Stars\altaffilmark{1}}

\author{Brian E. Wood}
\affil{Naval Research Laboratory, Space Science Division,
  Washington, DC 20375, USA; brian.wood@nrl.navy.mil}

\author{Hans-Reinhard M\"{u}ller}
\affil{Department of Physics and Astronomy, Dartmouth College,
  Hanover, NH 03755, USA}

\author{Graham M. Harper}
\affil{CASA, University of Colorado, Boulder, CO 80309-0389, USA}

\altaffiltext{1}{Based on observations made with the NASA/ESA Hubble
  Space Telescope, obtained at the Space Telescope Science Institute,
  which is operated by the Association of Universities for Research
  in Astronomy, Inc., under NASA contract NAS 5-26555.  These
  observations are associated with program GO-13462.  This paper
  also presents observations obtained with the Harlan J.\ Smith
  Telescope at McDonald Observatory of the
  University of Texas at Austin.}

\begin{abstract}

     We report on an ultraviolet spectroscopic survey of red
giants observed by the {\em Hubble Space Telescope}, focusing on
spectra of the Mg~II h \& k lines near 2800~\AA\ in order to study
stellar chromospheric emission, winds, and astrospheric absorption.
We focus on spectral types between K2~III and M5~III, a spectral type
range with stars that are noncoronal, but possessing strong,
chromospheric winds.  We find a very tight relation between Mg~II
surface flux and photospheric temperature, supporting the notion that
all K2-M5~III stars are emitting at a basal flux level.  Wind
velocities ($V_w$) are generally found to decrease with spectral type,
with $V_w$ decreasing from $\sim 40$ km~s$^{-1}$ at K2~III to $\sim
20$ km~s$^{-1}$ at M5~III.  We find two new detections of astrospheric
absorption, for $\sigma$~Pup (K5~III) and $\gamma$~Eri (M1~III).  This
absorption signature had previously only been detected for
$\alpha$~Tau (K5~III).   For the three astrospheric detections the
temperature of the wind after the termination shock correlates
with $V_w$, but is lower than predicted by the Rankine-Hugoniot shock
jump conditions, consistent with the idea that red giant termination
shocks are radiative shocks rather than simple hydrodynamic shocks.  A
full hydrodynamic simulation of the $\gamma$~Eri astrosphere is
provided to explore this further.

\end{abstract}


\section{Introduction}

     Ultraviolet and X-ray observations have demonstrated that among
main sequence stars stellar coronae are a ubiquitous phenomenon, with
coronal emissions being detectable for all stars with spectral types
later than A5~V \citep[e.g.,][]{js97}.  This is not the case, however, above
the main sequence.  For giant stars, F and G type giants are coronal,
but this coronal emission either disappears entirely or at least
dramatically weakens for K and M type giants.  For the K and M giants,
the coronae seem to be largely replaced by strong winds
($\dot{M}\sim 10^{-11}$ M$_{\odot}$~yr$^{-1}$) at chromospheric
temperatures ($T\sim 10^4$~K).  The first recognition of the existence
of this coronal dividing line among evolved stars between coronal and
``windy'' stars was by \citet{jll79}.  If attention is
limited to giant stars, i.e.\ luminosity class III stars, the dividing
line lies at a spectral type of about K2~III \citep{bh91}.
The nature of the transition across this dividing line is a topic of
particular interest.  Some stars to the right of the dividing line,
referred to as ``hybrid chromosphere'' stars, seem to maintain some
level of coronal emission, with $\gamma$~Dra (K5~III) being perhaps
the best-studied luminosity class III example \citep{tra97}.
This has led to the proposition that magnetic loops with coronal
temperatures still persist beyond the dividing line, but the coronal
emissions from these loops are completely or mostly hidden by the
dense, thick chromospheres that have developed for such stars, from
which emanate the strong chromospheric winds \citep{tra03}.  In order
to further investigate the transition from coronae to chromospheric
winds, we here present a new survey of red giant stars with spectral
types later than K2~III, using UV spectra taken by the {\em Hubble
Space Telescope} (HST).

     We are not only interested in
the winds of these stars, but also in the interactions of these winds
with the interstellar medium (ISM).  Ultraviolet spectroscopy from
HST has provided the first
spectroscopic detections of the wind-ISM interaction regions, or
``astrospheres,'' of many cool stars, including the heliospheric
structure surrounding our own Sun.  Hydrogen Lyman-$\alpha$ absorption
has been observed from the ``hydrogen wall'' region outside the
heliopause \citep{jll96}, and from the heliotail
\citep{bew14a}.  Hydrogen wall absorption has also been observed
around other Sun-like stars, representing the first method by which
solar-like coronal winds can be detected around other stars
\citep{bew05a,bew14b}.

    However, we are here interested in the astrospheres of red giant
stars.  Observations of the Mg~II h \& k
lines near 2800~\AA\ have been used to detect absorption from the
astrosphere of $\alpha$~Tau (K5~III)
\citep{rdr98,bew07}.  Models of the $\alpha$~Tau
astrosphere were computed using codes designed to study the heliosphere.
Analogous to the solar wind, the red giant wind expands radially from the
star until it reaches the termination shock (TS), where it is shocked
to subsonic speeds.  The post-TS wind is slower, hotter, and more
dense.   It is this post-TS region that provides Mg~II column densities
sufficiently high to yield the observed Mg~II astrospheric
absorption.

    Due to its high positive stellar radial velocity
($V_{rad}=+54.3$ km~s$^{-1}$), our line of sight towards $\alpha$~Tau
is estimated to be at an angle of $\theta\sim 170^{\circ}$ from the
upwind direction of the ISM flow vector in
the rest frame of the star.  This line of sight is therefore very much
through the ``astrotail,'' and this may be essential for detecting
the astrospheric absorption.  A downwind line of sight provides a much
longer path length through the post-TS material than an upwind or
sidewind line of sight, leading to much higher column densities through
the post-TS material \citep{bew07}.  Nevertheless, the
hydrodynamic models of the $\alpha$~Tau astrosphere were unable to
reproduce the observed astrospheric absorption very well, predicting
far too little absorption, and placing the absorption farther from
the stellar rest frame than observed.  The proposed explanation was
that the $\alpha$~Tau TS is a radiative shock, with radiative cooling
from H Lyman lines yielding further compression and deceleration
behind the TS, resulting in a stronger Mg~II absorption feature closer
to the rest frame of the star, in better agreement with the observed
absorption \citep{bew07}.

     The $\alpha$~Tau observation has long been the only example
of a detected red giant astrosphere.  Our new HST survey of red giant
stars was not only designed to study the red giant winds, but also
designed to try to find new detections of astrospheric absorption
around such stars.

\section{The HST Red Giant Survey}

     We are interested in the warm, partially ionized winds and
resulting astrospheres of giants with spectral types between K2~III
and M5~III.  We do not consider stars later than M5,
because such giants contain Miras and other long period variables,
which have cooler and more massive winds of a different character,
and we do not consider giants earlier than K2, which lie on the
wrong side of the coronal dividing line mentioned in section~1,
exhibiting high-temperature coronal emission instead of
strong chromospheric winds \citep[e.g.,][]{bh91}.  Based on
the idea that a downwind line of sight through the astrotail is
essential for detecting the Mg~II astrospheric absorption signature,
all the selected targets have $V_{rad}\geq +40$ km~s$^{-1}$.  This
choice has the added benefit of shifting the astrospheric absorption
and stellar wind absorption away from the expected locations of narrow
interstellar Mg~II absorption, which can confuse analysis of the line
profile.

\begin{table}[t]
\scriptsize
\begin{center}
\caption{HST red giant targets}
\begin{tabular}{lcccccccccccc} \hline \hline
Star & Alternate & Spect.\ & V$^a$ & V-K$^b$ & T$_{eff}^c$ &    R$^d$  & d$^e$ &
  $V_{rad}$$^f$ & log F$_{\rm Mg II}$$^g$ & $V_w$$^h$ &$V_{ISM}$$^i$ & $\theta^j$ \\
     &   Name    & Type    &   &     &(K)&(R$_{\odot}$) & (pc)   & (km/s) &
                      & (km/s) & (km/s)  & (deg) \\
\hline
\multicolumn{13}{l}{\underline{New Survey Targets}} \\
HD 66141  & HR 3145      & K2 III   & 4.38 & 2.97 & 4274 & 23.9 & 77.9 & 71.6 &
  5.08 &   ...   & 76 & 142 \\
HD 211416 & $\alpha$~Tuc & K3 III   & 2.82 & 2.91 & 4310 & 37.3 & 61.2 & 42.2 &
  5.02 &   ...   & 52 & 140 \\
HD 87837  & HR 3980      & K4 III   & 4.38 & 3.34 & 4086 & 33.6 & 90.6 & 39.8 &
  4.69 &   ...   & 45 & 143 \\
HD 50778  & $\theta$~CMa & K4 III   & 4.08 & 3.42 & 4052 & 35.4 & 79.9 & 96.2 &
  4.96 & $30\pm5$& 95 & 150 \\
HD 59717  & $\sigma$~Pup & K5 III   & 3.25 & 3.66 & 3959 & 43.7 & 59.4 & 87.3 &
  4.84 & $43\pm7$& 88 & 147 \\
HD 25025  & $\gamma$~Eri & M1 III   & 2.94 & 3.87 & 3889 & 58.9 & 62.3 & 60.8 &
  4.67 & $24\pm7$& 57 & 146 \\ 
HD 44478  & $\mu$~Gem    & M3 III   & 2.87 & 4.73 & 3675 &107.7 & 71.0 & 54.4 &
  4.50 & $19\pm5$& 55  & 144 \\
HD 20720  & $\tau^4$~Eri & M3.5 III & 3.70 & 4.85 & 3652 &102.9 & 93.4 & 41.7 &
  4.49 & $23\pm7$& 34 & 149 \\
HD 120323 & HR 5192      & M4.5 III & 4.19 & 5.85 & 3500 & 82.4 & 56.1 & 40.7 &
  4.20 & $19\pm5$& 43 & 172 \\
\multicolumn{13}{l}{\underline{Archival Stars}} \\
HD 124897 & $\alpha$~Boo & K2 III   &$-0.05$&2.86 & 4341 & 25.2 & 11.3 &$-5.2$&
  5.23 & $43\pm4$&113 &  92 \\
HD 164058 & $\gamma$~Dra & K5 III   & 2.23 & 3.58 & 3989 & 53.4 & 47.3&$-27.9$&
  4.79 &   (74)  & 18 &  40 \\
HD 29139  & $\alpha$~Tau & K5 III   & 0.86 & 3.90 & 3880 & 51.0 & 20.4 & 54.3 &
  4.77 & $35\pm5$& 44 & 170 \\
HD 108903 & $\gamma$~Cru & M3.5 III & 1.64 & 4.76 & 3669 & 73.9 & 27.2 & 21.0 &
  4.49 & $28\pm9$& 37 & 116 \\
\hline
\end{tabular}
\end{center}
\tablecomments{$^a$V band magnitude. $^b$V-K
  color. $^c$Photospheric effective temperature. $^d$Stellar
  radius. $^e$Stellar distance. $^f$Heliocentric radial
  velocity. $^g$Logarithmic Mg~II h \& k surface flux (in
  ergs~cm$^{-2}$~s$^{-1}$ units). $^h$Stellar wind velocity measured
  from the Mg~II k line. $^i$Interstellar wind velocity seen by the
  star in the stellar rest frame. $^j$Angle between the upwind
  direction of the ISM flow vector in the rest frame of the star and
  our line of sight to the star.}
\normalsize
\end{table}
     Table~1 lists properties of the nine red giants that constitute
our new red giant survey sample.  Within the HST archives, there are
high resolution Mg~II spectra of four other K2-M5~III stars.  These
stars are also listed in Table~1, as we will consider their Mg~II
spectra as well, although only $\alpha$~Tau has the high
$V_{rad}$ that we presume to be advantageous for astrospheric
detection.  The properties listed in Table~1 are spectral type, V
band magnitude, V-K color, temperature, radius, distance, radial
velocity ($V_{rad}$), logarithmic Mg~II surface flux (F$_{\rm Mg II}$,
in ergs~cm$^{-2}$~s$^{-1}$ units), stellar wind terminal velocity
($V_w$), the ISM velocity seen by the star ($V_{ISM}$), and the
orientation of our line of sight to the star with respect to the
upwind direction ($\theta$).  The SIMBAD database was utilized to find
the V magnitudes, V-K colors, distances \citep{fvl07}, and
heliocentric radial velocities.  The V-K colors are used in the
estimation of both the photospheric effective temperatures and stellar
radii, based on the prescriptions of \citet{ab10} and \citet{gtvb99},
respectively.  The radii computed in this manner should have
uncertainties of 11.7\% \citep{gtvb99}.  Five of the stars have
interferometric radii measured by \citet{dm03}.  The only one
initially found to be in poor agreement was $\mu$~Gem.  The
discrepancy was resolved by replacing the initially assumed magnitude
with the 2MASS catalog's $K=-1.86$ value \citep{rmc03}, changing the
V-K color and inferred radius significantly.

     In computing the last two quantities of Table~1 regarding the
ISM flow direction and magnitude at each star, which are relevant for
astrospheric studies, we note that our target stars are all close
enough to be within the Local Bubble (LB), a region of generally low
ISM density extending roughly 100~pc from the Sun in most directions
\citep{dms99,rl03}.  We verify that the stars are within the LB
using the LB maps of \citet{rl03}.  Most of the volume of
the LB is believed to be very hot, ionized plasma, which produces much
of the soft X-ray background emission \citep{sls98,sls14}.
The flow vector of the LB material is uncertain, but a plausible
assumption is that it is roughly at rest relative to the Local
Standard of Rest (LSR) \citep[e.g.,][]{wd98}.  Combining this
assumption with the known proper motion and radial velocity of our
target stars, we can estimate the ISM flow vector seen in the rest
frame of the star and its orientation relative to our line of sight to
the star, characterized by the $V_{ISM}$ and $\theta$ parameters in
Table~1.  All the new survey stars have $\theta\geq 140^{\circ}$,
confirming that the lines of sight to these stars are expected to be
very downwind, which was one of the desired selection criteria to
maximize the likelihood of detecting astrospheric absorption.

     The nine newly observed stars were observed by the Space
Telescope Imaging Spectrograph (STIS) instrument on HST
\citep{rak98,bew98}, with the observations occurring
between 2013 October and 2015 January.  For each of our targets, the
HST visit consisted of a spectrum of the 2574--2851~\AA\ wavelength
range taken with STIS's high resolution ($R\equiv
\lambda/\Delta\lambda = 110,000$) E230H grating, followed by a longer
exposure of the 1150--1700~\AA\ wavelength range taken with the
moderate resolution ($R=46,000$) E140M grating.  Our focus here is on
the Mg~II h \& k lines within the E230H spectrum, with rest
wavelengths of 2803.5315~\AA\ and 2796.3543~\AA, respectively
\citep{dcm03}. The $\alpha$~Tau spectrum considered is the same as that
studied by \citet{bew07}, which is from 1994~April~8, and is a
product of the Ech-B grating on the UV spectrometer that preceded STIS
on HST, the Goddard High Resolution Spectrograph (GHRS).  The other
three archival Mg~II spectra considered are all STIS/E230H
observations, from 1998~August~24, 2011~August~11, and 2011~August~4
for $\alpha$~Boo, $\gamma$~Dra, and $\gamma$~Cru, respectively.
\begin{figure}[p]
\plotfiddle{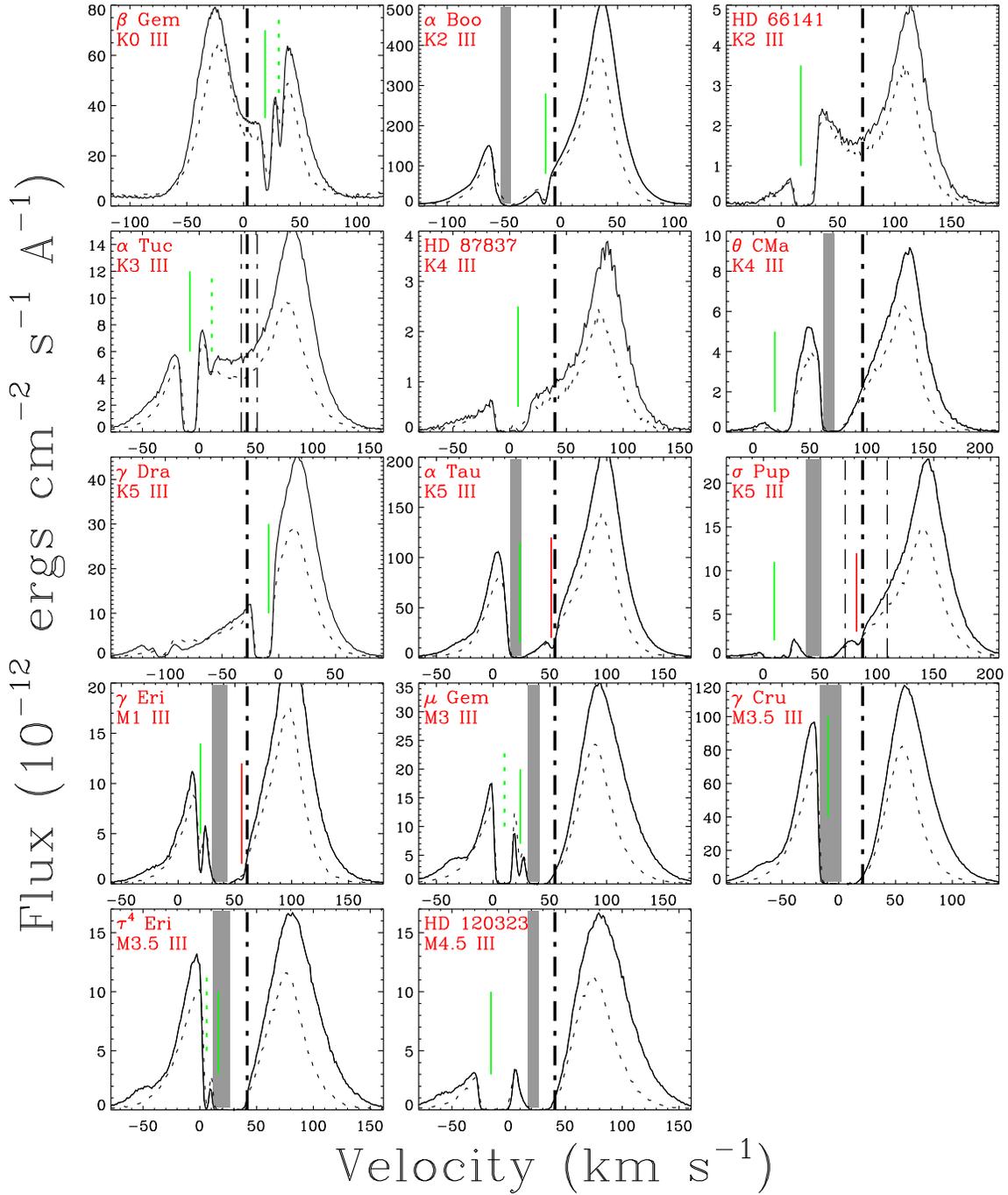}{7.4in}{0}{100}{100}{-305}{-130}
\caption{Mg~II h (dotted) and k (solid) lines observed by HST are
  plotted on a heliocentric velocity scale, with the stars shown in
  order of spectral type.  Although not actually in our sample of
  K2-M5~III stars, $\beta$~Gem (K0~III) is included here to illustrate what
  Mg~II profiles look like for a coronal giant without a strong wind.
  Thick vertical dot-dashed lines indicate the stellar rest frame
  ($V_{rad}$), although for two stars in spectroscopic binary systems
  ($\alpha$~Tuc and $\sigma$~Pup) it is actually the systemic
  center-of-mass velocity, and the thin dot-dashed lines around it
  mark the range of orbital motion of the red giant.  Green lines
  indicate ISM absorption features, with solid lines marking the
  expected location of LIC absorption, and dotted lines identifying
  other absorption that we believe is from the ISM.  Shaded regions
  indicate the estimated stellar wind terminal velocity and its
  uncertainty, for stars with deep, opaque wind absorption.  Red lines
  mark astrospheric absorption; only seen for $\alpha$~Tau,
  $\sigma$~Pup, and $\gamma$~Eri.}
\end{figure}

     Figure~1 displays the Mg~II spectra of all the K2-M5~III stars
listed in Table~1, in spectral type order.  We also include in the
figure an archival HST/GHRS Mg~II spectrum of the K0~III star
$\beta$~Gem \citep{ard97}.  Being on the coronal side of the
dividing line mentioned in section~1, the $\beta$~Gem spectrum serves
to illustrate what a red giant chromospheric Mg~II line profile should
look like in the absence of an opaque stellar wind.  The stellar
rest frame is indicated in Figure~1, but for two stars in
our sample that are spectroscopic binaries the figure instead
indicates the systemic center-of-mass radial velocity and the
range of velocities covered by the red giants in their elliptical
orbits.  These two systems are $\alpha$~Tuc (K3~III+?) and
$\sigma$~Pup (K5~III+G5~V), with orbital periods of 4197.7 and
257.8 days, respectively \citep{dp04}.  The actual
radial velocities of the red giants at the times of observation are
best estimated from narrow H$_2$ lines in the FUV E140M spectrum,
which suggest radial velocities of 42 and 103 km~s$^{-1}$ for
$\alpha$~Tuc and $\sigma$~Pup, respectively.  The implications of
$\sigma$~Pup's binarity will be discussed in detail in section~6.

     The Mg~II profiles are complicated by the presence of intervening
absorbers in between the stellar chromosphere and the observer.  These
absorbers include the stellar wind, the ISM, and the stellar astrosphere.
Given that all our stars are relatively nearby, we expect the ISM
absorption to be relatively close to the velocity predicted by the local
interstellar cloud (LIC) vector measured for the ISM cloud immediately
surrounding the Sun \citep[e.g.,][]{sr08,bew15}.
Multiple ISM velocity components are often observed even
for very nearby stars, but not typically at velocities separated by
more than $\sim 20$ km~s$^{-1}$ from the LIC velocity \citep{sr02}.

     The LIC velocity is shown explicitly in all the Figure~1
panels.  There is always absorption observed at that location.
Although the figure identifies the absorption as LIC, there could be a
blend of ISM velocity components, particularly in cases like HD~87837
where the absorption feature is particularly broad and not well
centered on the LIC velocity.  In several cases ($\alpha$~Tuc,
$\mu$~Gem, and $\tau^4$~Eri) there is a separate, distinct narrow
absorption feature adjacent to the LIC absorption that we suspect is
interstellar, and it is identified as such in Figure~1.  Our selection
of stars with high, positive radial velocities shifts the stellar wind
absorption redward of the ISM absorption in all cases, generally far
enough to adequately separate the two.  Among our new survey targets,
only for $\tau^4$~Eri have we ended up with a case where the LIC
absorption is clearly within the saturated core of the wind
absorption, making it impossible to separate the two and complicating
analysis of the absorption.

     We compute integrated Mg~II h and k line fluxes for all the red
giant target stars.  We correct for ISM absorption by interpolating a
continuum over the ISM features identified in Figure~1.  Figure~2a
shows the k/h flux ratio plotted versus photospheric temperature.  The
ratio clearly increases towards later spectral types.  The ratio would
be 2 in an optically thin plasma, but stellar chromospheres are highly
optically thick to Mg~II emission, so values less than 2 are always
observed \citep{gsb79,cb82,mcs92}.
The spectral type dependence of k/h seen in Figure~2a,
which is not seen for main sequence stars, is indicative of systematic
changes in the chromospheric opacity towards the later type giants,
which will have lower surface gravities due to larger radii (see
Table~1).  Past solar and stellar chromospheric models have
discussed the diagnostic potential of the k/h ratio
\citep[e.g.,][]{jll70,jl13}.

     After summing up the flux of both the h and k lines, we can
compare the HST fluxes with those measured previously by the
{\em International Ultraviolet Explorer} (IUE).  For the eight stars
we have in common with the \citet{mipm11} compilation of evolved star
IUE Mg~II fluxes we find reasonably good agreement, with no systematic
flux discrepancies, and an average difference of only 20.8\%.  We
convert the fluxes at Earth to surface fluxes (F$_{\rm Mg II}$) in
ergs~cm$^{-2}$~s$^{-1}$ units, which are listed logarithmically in
Table~1.  The biggest uncertainty in the surface fluxes lies in the
stellar radii.  The 11.7\% angular diameter uncertainty from
\citet{gtvb99} translates into a 23.4\% error in F$_{\rm Mg II}$,
which leads to 0.09~dex uncertainties for the $\log$~F$_{\rm Mg II}$
values in Table~1.

     In Figure~2b, the Mg~II fluxes are plotted versus temperature.
There is a strong correlation, and a simple line fits the data
quite well, suggesting
\begin{equation}
\log F_{\rm Mg II}=9.109\log T_{eff} - 28.000.
\end{equation}
The correlation is apparent whether F$_{\rm Mg II}$ is plotted
versus spectral type, V-K, or temperature.
The 1$\sigma$ scatter about the fitted line is only 0.054 dex, or
about 13\%, which is less than the 23.4\% F$_{\rm Mg II}$ uncertainty
estimated above.  The remarkably tight correlation between
chromospheric activity and temperature (or spectral type) seen
for red giants is very different from what is found for main
sequence stars.  For main sequence stars, the Mg~II fluxes span
a range of about a factor of 30 at a given spectral type,
due to chromospheric and coronal emission being
largely determined by rotation rate, which can vary greatly
\citep[e.g.,][]{mm95,ir05,bew05b}.
\begin{figure}[t]
\plotfiddle{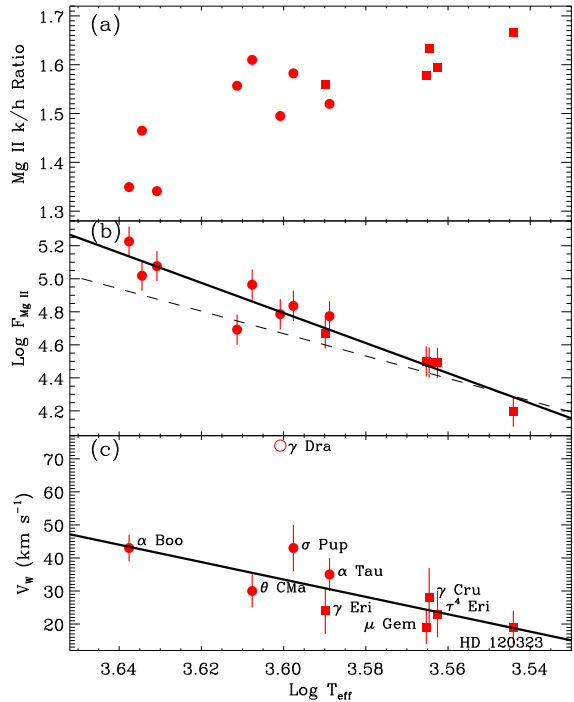}{3.4in}{0}{50}{50}{-160}{-65}
\caption{(a) Mg~II k/h flux ratio plotted versus  photospheric
  temperature for K (circles) and M (squares) giants. (b) Mg~II surface
  fluxes versus photospheric temperature, with a linear fit.
  The dashed line is the ``V-K Model'' basal flux line of
  P\'{e}rez Mart\'{i}nez et al.\ (2011).  (c) Stellar wind velocity
  versus temperature, with a linear fit that excludes the discrepant
  $\gamma$~Dra.}
\end{figure}

     The homogeneity of K2-M5~III Mg~II fluxes at a given spectral
type is consistent with the idea that on the noncoronal side of
the dividing line giant stars are all rotating very slowly, and at
a given spectral type the stars are therefore
producing chromospheric emission at basically the same basal
flux level \citep{cjs87,mipm11}.
We found rotational velocity measurements for eight of our thirteen
stars \citep{jrd99,jrd14},
and all are slow rotators.  Only $\sigma$~Pup has a rotational
velocity above 2 km~s$^{-1}$, with $v\sin i=2.9\pm 1.1$ km~s$^{-1}$.
In Figure~2b, our Mg~II fluxes are compared with the ``V-K Model''
basal flux line of \citet{mipm11}, based on IUE
measurements of many more stars over a wider range of spectral
types, albeit with many coronally active stars not emitting at a basal
level.  There is reasonably good agreement, though our measurements
suggest a slightly steeper relation, $F_{\rm MgII}\propto T_{eff}^{9.1}$
instead of $F_{\rm MgII}\propto T_{eff}^{6.8}$, at least in this spectral
type range.

     One paradigm for the basal chromosphere phenomenon is that it
represents the chromospheric heating level due to the dissipation
of acoustic waves generated from photospheric oscillations, and in the
absence of significant magnetic activity this is the dominant
chromospheric heating mode \citep{bb98,wr03,pu05,mc07}.  Acoustic
heating models appear to be able to successfully reproduce the
observed dependence of basal flux on photospheric temperature, a
dependence that is similar on both giants and main sequence stars.
However, alternative views exist, which suggest that a base level of
magnetic activity is instead responsible \citep[e.g.,][]{pgj98}.
Support for this view comes from the existence of a basal
level of coronal X-ray emission for main sequence stars, which it is
widely accepted requires a magnetic origin \citep{ks89,js97,kps12}.
It is less clear that
a basal X-ray emission level exists for K2-M5~III stars, as so few
have reliable X-ray detections \citep{tra03}.  Nevertheless,
the detection of magnetic fields on the supergiant Betelgeuse
\citep{ma10} and the Mira $\chi$~Cyg \citep{al14},
as well as the observation of an X-ray outburst on Mira
itself \citep{mk05}, demonstrates that even very cool,
evolved, and slowly rotating stars still have some magnetic activity,
and many models of red giant chromospheres and wind acceleration rely
on the presence of magnetic fields and Alfv\'{e}n waves
\citep{tks07,va10}.

\section{The Stellar Wind Absorption Profiles}

     Even in the absence of the ISM absorption, wind scattering,
and astrospheric
absorption, the Mg~II line profiles are already complex due to the
high opacity of the chromospheric Mg~II lines.  The $\beta$~Gem line
profile in Figure~1 is double-peaked, with a stronger blue peak.  Such
``self-reversals'' are typical for chromospheric Mg~II lines,
including ones observed from the Sun, and the blue peak is commonly
found to be the stronger one \citep{rfd94}.  None of
the K2-M5~III Mg~II profiles are like that of $\beta$~Gem.  The
closest are the spectra of HD~66141 (K2~III) and $\alpha$~Tuc
(K3~III), which might be interpreted as being chromospheric lines that
happen to have stronger red peaks, but it is far more likely that this
shift in flux from blue peak to red peak is an indication of a
chromospheric wind scattering flux from the blue side of the line to
the red.

     The wind will naturally be most opaque just bluewards of the
stellar rest frame, extending to the wind's terminal velocity, and even
beyond it to an extent determined by the turbulent velocities in the
wind at its terminal speed.  The wind will be a photon-scattering
environment in which Mg~II photons absorbed and reemitted by Mg$^+$
ions in the wind will preferentially escape by frequency scattering
away from the velocities at which the wind is most opaque
\citep[e.g.,][]{gmh95}.  For most of the stars in Figure~1,
the result is a broad, deep, saturated wind absorption feature
blueward of the stellar rest frame.

     However, for three of the stars close to the dividing line at
K2-K4~III (HD~66141, $\alpha$~Tuc, and HD~87837) the wind is
apparently weaker and less opaque, and the effect on the line profile
is more subtle, yielding merely a line asymmetry with more flux
emerging redward of the stellar rest frame.  The exception is
the well-studied K2~III star $\alpha$~Boo, which shows deep wind
absorption despite being very close to the coronal dividing line,
implying a high mass loss rate of
$\dot{M}=2\times 10^{-10}$ M$_{\odot}$~yr$^{-1}$ \citep{eog13}.

     For the stars that have deep wind absorption, we can estimate
the terminal wind speed ($V_w$) empirically, based on the following
prescription.  First, we determine the maximum flux level within the
Mg~II k line blueward of the stellar rest frame (e.g.,
$\sim 5\times 10^{-12}$ ergs~cm$^{-2}$~s$^{-1}$~\AA$^{-1}$ for
$\theta$~CMa).  Second, we divide this value by an admittedly arbitrary
factor of 20, and we then determine at what velocity that flux level
is reached on the blue side of the wind absorption ($V_b$).  In the
stellar rest frame, this represents an upper bound for $V_w$.  Next,
we determine at what velocity that flux level is reached on the red
side of the wind absorption ($V_r$).  The average of $V_b$ and $V_r$
represents an estimate of the center of the wind absorption, which in
the stellar rest frame provides an estimate of the lower bound of
$V_w$.  The shaded regions in Figure~1 indicate the range of $V_w$
implied by these lower and upper bounds for all the stars.  If
microturbulent velocities in the stellar wind at the terminal velocity
are low (e.g., a few km~s$^{-1}$) then the upper bound (i.e., the left
edge of the shaded region) should be close to $V_w$; but if turbulent
velocities are high then the lower bound (i.e., the right edge of the
shaded region) would be a better estimate \citep{gmh95}.  In
any case, with the bounds on $V_w$ now defined, we can now quote an
estimate of $V_w$ with error bars, which are listed in Table~1.
Mathematically, if V$_b$ and V$_r$ are in a heliocentric rest frame,
as in Figure~1, then $V_w=V_{rad}-[V_b+(V_r-V_b)/4]$, with uncertainty
$\Delta V_w=(V_r-V_b)/4$.

     We cannot use this prescription for the first three giants
listed in Table~1, which do not have the deep, saturated wind
absorption features (see Figure~1).  The K5~III giant $\gamma$~Dra
is a special case.  It does not possess a broad, deep stellar wind
absorption feature, but there is a rather narrow dip at a
heliocentric velocity of $-102$ km~s$^{-1}$ in Figure~1, which
is likely indicative of the wind terminal velocity.  This
implies a very high wind speed of $V_w=74$ km~s$^{-1}$.  An older
HST/GHRS spectrum of $\gamma$~Dra studied by \citet{rdr98}
shows a similar profile, but with the absorption feature less
blueshifted, suggesting $V_w=67$ km~s$^{-1}$.  However,
\citet{rdr98} find it difficult to interpret the
Mg~II profile with a single wind component, and propose the
existence of a slower wind component (with $V_w=30$ km~s$^{-1}$)
in addition to the fast one
producing the narrow absorption dip.  The speed of this
slower wind component is more consistent with that observed
for the other K giants.

     The Mg~II profiles of $\gamma$~Dra are in most respects
similar to those of HD~87837 (K4~III), but HD~87837 does not have the
absorption dip far from the stellar rest frame that would suggest that
there is necessarily opaque stellar wind material at those high
speeds.  It is tempting to associate $\gamma$~Dra's unusually fast
wind with the star also being one of the most well-known
hybrid-chromosphere giants, with weak but detectable coronal X-ray
emission \citep{tra06}, and detectable transition region UV
emission from lines such as C~IV $\lambda$1548 \citep{tra97}.
\citet{gmh95} found a very high wind speed of $V_w=104$
km~s$^{-1}$ for $\alpha$~TrA (K4~II), which is another canonical
hybrid chromosphere star.  Before the HST era, \citet{dr82}
previously noted a possible correlation between the hybrid
chromosphere stars and high speed Mg~II wind absorption features in
IUE spectra.  Thus, both the HST and IUE databases of red giant
spectra seem to be consistent with the notion that red giants with
unusually strong transition region and coronal emissions also have
high speed wind components.

     The wind velocities of the M giants are clearly lower than
those of the K giants in our sample, with $V_w\sim 40$ km~s$^{-1}$
for K2~III stars near the coronal dividing line, and
$V_w\sim 20$ km~s$^{-1}$ at M5~III.  The correlation is
shown explicitly in Figure~2c, which plots $V_w$ versus photospheric
temperature. Excluding the clearly discrepant $\gamma$~Dra, the
linear fit is
\begin{equation}
V_w=263.16\log T_{eff}-913.90.
\end{equation}
This is the clearest demonstration of such a correlation for red
giants, though past Mg~II observations from IUE have also suggested a
trend of this nature \citep{sad86a,sad86b,pgj91}.

     There is some ambiguity with regards to whether the
optical spectra of Ca~II H \& K wind absorption support a wind
velocity correlation with spectral type.  The wind velocities quoted
by \citet{dr77} would seem to suggest lower velocities for M giants,
but inferences of terminal velocities from the Ca~II data are open to
interpretation.  At least within the M giants alone, which are the
only class III stars with reliably deep, saturated wind absorption,
the centroid velocity of the Ca~II absorption decreases towards later
spectral types but the blueward extent of the absorption does not,
leading to the early interpretation that these variations are
indicative of changes in line formation depth and turbulent velocity
with spectral type rather than any change in wind velocity
\citep{ajd60,dr75,amb79}.  The very presence of Ca~II wind absorption
implies that the stellar wind must be relatively cold close to the
star.

     We can further explore the comparison of Ca~II and Mg~II
wind diagnostics using more recent Ca~II observations, specifically
spectra taken at a very high resolution of $R=200,000$ with the
Coude cross-dispersed spectrometer (cs21) on the 2.7-meter
Harlan J.\ Smith Telecope at McDonald Observatory.
In 2006 January, eight K0-M5 evolved stars were
observed:  $\beta$~Gem (K0~III), $\alpha$~Boo (K2~III),
$\beta$~UMi (K4~III), $\alpha$~Tau
(K5~III), $\beta$~And (M0~III), $\alpha$~Cet (M1.5~III), $\beta$~Peg
(M2.5 II-III), and $\alpha$~Her~A (M5~II).  The spectral
type range covered resembles our HST red giant sample, although the
last two stars drift out of luminosity class III.  Figure~3 shows the
resulting spectra of the Ca~II K line.
\begin{figure}[t]
\plotfiddle{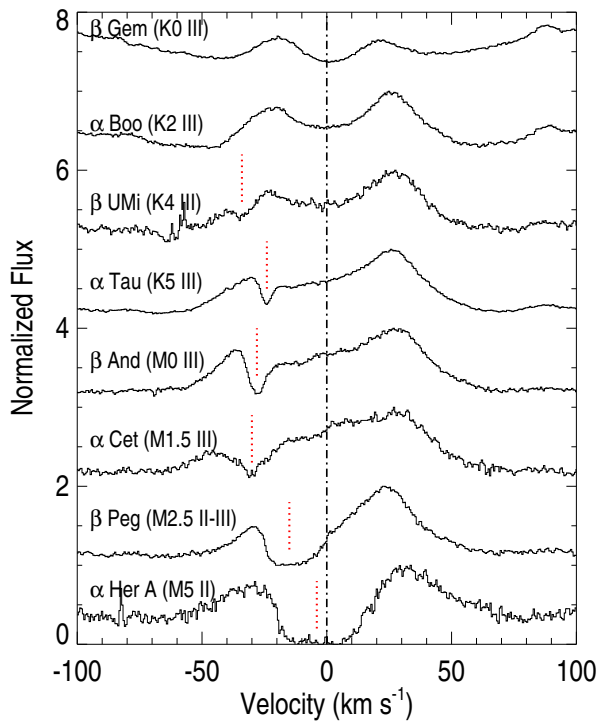}{3.5in}{90}{55}{55}{220}{-35}
\caption{Ca~II K line profiles of 8 evolved stars, observed by the
  Harlan J.\ Smith Telecope at McDonald Observatory.  The spectra
  are shown on a velocity scale centered on the rest frame of the
  star (dot-dashed line).  For all stars except $\beta$~Gem, wind
  scattering leads to a red peak stronger than the blue peak,
  and in most cases to clear absorption features marked with
  dotted lines.}
\end{figure}

     Analogous to Figure~1, $\beta$~Gem is used in Figure~3 to
indicate what the Ca~II line profile looks like for a coronal star
without a strong chromospheric wind, and as for Mg~II, the line
profile is double-peaked, with a stronger blue peak.
All of the other Ca~II K lines in Figure~3 have stronger red peaks,
suggesting the presence of a wind.
However, the latest type stars, $\beta$~Peg and $\alpha$~Her~A, are the
only ones that exhibit deep, saturated Ca~II wind absorption
profiles.  The wind signatures are more subtle for the earlier
type stars.  The $\alpha$~Boo and
$\beta$~UMi Ca~II profiles are very similar to the Mg~II profiles of
HD~66141 and $\alpha$~Tuc (see Figure~1).  It is interesting that
$\alpha$~Boo, which has strong and deep Mg~II wind absorption in
Figure~1, does not exhibit analogously strong Ca~II absorption in
Figure~3.  Several of the stars ($\alpha$~Tau, $\beta$~And,
$\alpha$~Cet, and perhaps $\beta$~UMi) show relatively narrow, weak
Ca~II absorption between $-34$ and $-24$ km~s$^{-1}$ that should be
indicative of the stellar wind's terminal speed.  These features
resemble the narrow Mg~II absorption seen in Figure~1 for
$\gamma$~Dra, albeit not at nearly as high a velocity.  The $-34$ to
$-24$ km~s$^{-1}$ range of velocities is comparable to the wind speeds
inferred from Mg~II.

     However, it is worth noting that for $\alpha$~Tau, we inferred
$V_w=35\pm 5$ km~s$^{-1}$ from Mg~II, while the $\alpha$~Tau Ca~II
absorption in Figure~3 would suggest a lower value of $V_w\approx 24$
km~s$^{-1}$.  This is likely due to actual variability in the wind
of $\alpha$~Tau.  The variability of the Ca~II wind absorption seen
from $\alpha$~Tau has long been established \citep{dr77,wlk78}.  More
recently, \citet{gmh11} found the wind velocity inferred from
Ca~II wind absorption to increase from 20.9 to 28.0 km~s$^{-1}$
between 2005 September and 2007 March.

\section{New Astrospheric Absorption Detections}

     Based on the $\alpha$~Tau example, astrospheric Mg~II
absorption should be weak, narrow, and slightly blueshifted from the
stellar rest frame.  Only two of the other stars in our sample,
$\sigma$~Pup and $\gamma$~Eri, show absorption of this nature (see
Figure~1).  We not only look for astrospheric absorption in the Mg~II
lines, but also in other lines within our HST spectra in which it
might be observed, particularly the O~I line at 1302.2~\AA, C~II at
1334.5~\AA, and Fe~II at 2600.2~\AA, which all have line profiles
similar to Mg~II, with the deep wind absorption trough.  For
$\gamma$~Eri we find a marginal detection of Fe~II absorption, while for
$\sigma$~Pup we find more convincing detections for both
Fe~II and C~II.  Given that for $\alpha$~Tau the absorption signature
was only seen in Mg~II, these are the first detections of the
astrospheric absorption signature in any line but Mg~II.  Figure~4
displays all astrospheric absorption lines detected to date, for all
three red giants with detected astrospheres.

\begin{table}[t]
\scriptsize
\begin{center}
\caption{Astrospheric Absorption Line Measurements}
\begin{tabular}{lcccccc} \hline \hline
Star & Species &   V$^a$  &  b$^b$   & log N$^c$ & $\eta^d$ & T$^e$ \\
     &         &(km/s)&(km/s)&       & &(10$^4$ K) \\
\hline
$\sigma$~Pup & Mg II & $83.7\pm 0.1$ & $3.88\pm 0.16$ & $12.064\pm 0.012$ &
  $11.9\pm 2.1$ & $2.19\pm 0.18$ \\
             & Fe II & $85.0\pm 0.8$ & $2.9\pm 1.5$   & $12.38\pm 0.17$ &
  ... & ... \\
             & C II  & $82.9\pm 1.8$ & $4.9\pm 3.1$   & $13.47\pm 0.20$ &
  ... & ... \\
$\gamma$~Eri & Mg II & $56.7\pm 0.1$ & $2.51\pm 0.13$ & $12.213\pm 0.013$ &
   $5.9\pm 1.7$ & $0.92\pm 0.09$ \\
             & Fe II & $56.1\pm 1.5$ & $1.1\pm 1.7$   & $12.15\pm 0.38$ &
  ... & ... \\
$\alpha$~Tau & Mg II & $52.3\pm 0.1$ & $3.55\pm 0.04$ & $12.230\pm 0.003$ &
  $17.5\pm 3.1$ & $1.83\pm 0.04$ \\
\hline
\end{tabular}
\end{center}
\tablecomments{$^a$Heliocentric velocity. $^b$Doppler broadening
  paramter. $^c$Logarithmic column density (in cm$^{-2}$ units).
  $^d$Termination shock compression ratio. $^e$Post termination shock
  temperature.}
\normalsize
\end{table}
     We have fitted these absorption features with single
absorption lines, analyzing them in the same fashion as
ISM absorption lines \citep[e.g.,][]{sr02,sr04}.  Required
oscillator absorption strengths are taken from \citet{dcm03}.  Each
absorption component is defined by three parameters: central velocity
($V$), Doppler broadening parameter ($b$), and column density ($N$).
To constrain the fit as much as possible, the two Mg~II lines are
fitted simultaneously, with self-consistent fit parameters.  The
resulting fits to the data are shown in Figure~4, taking into account
instrumental broadening using line spread functions from \citet{sh12}.
The fit parameters are listed in Table~2.  The column
densities, with units of cm$^{-2}$, are listed in logarithmic form.
The Fe~II and C~II data are noisy, and the C~II line is not well
resolved in the E140M spectrum.  As a consequence, uncertainties
in the Fe~II and C~II fit parameters are large.
\begin{figure}[t]
\plotfiddle{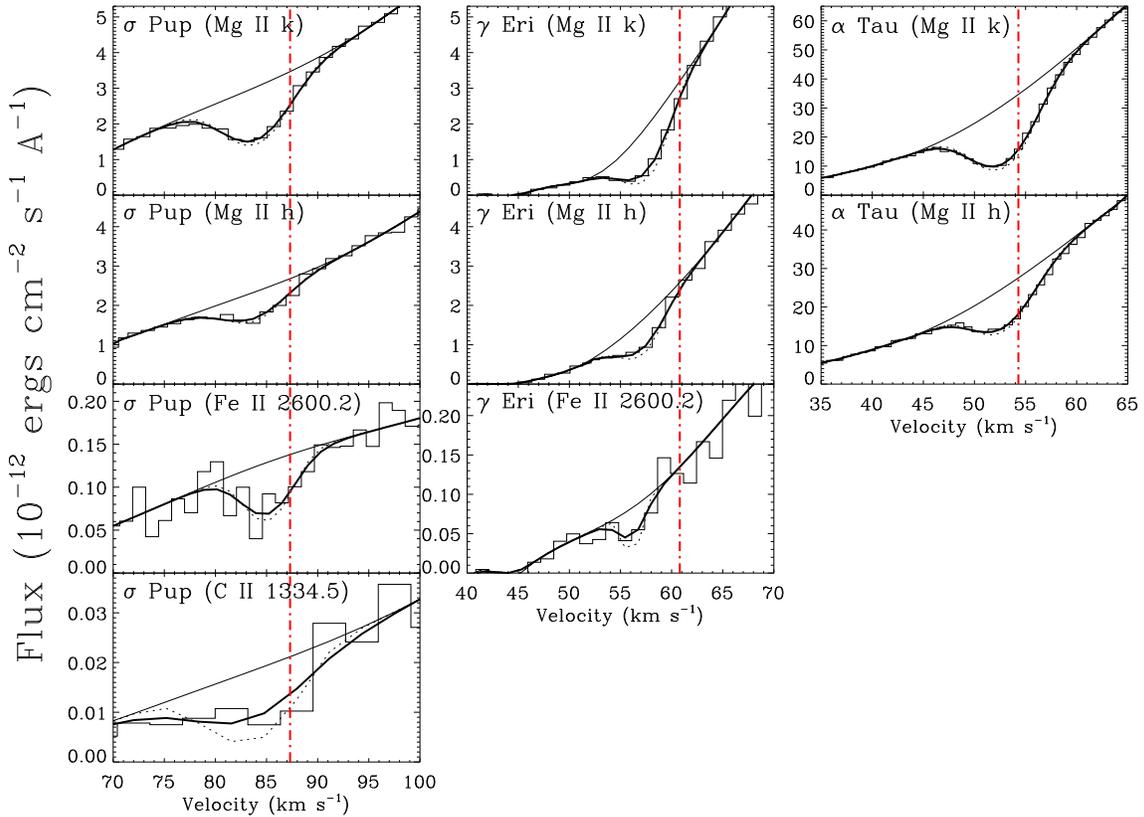}{4.0in}{90}{65}{65}{250}{-45}
\caption{Fits to astrospheric absorption lines observed towards
  $\sigma$~Pup (left), $\gamma$~Eri (middle), and $\alpha$~Tau
  (right), plotted on a heliocentric velocity scale.  Dotted and
  solid lines show the fits before and after convolution with the
  instrumental line profiles, respectively.  Vertical dot-dashed
  lines indicate the stellar rest frame.}
\end{figure}

     The velocities of the astrospheric absorption listed in
Table~2 indicate the post-TS velocity, after being subtracted from
$V_{rad}$ to place them in the proper stellar rest frame.  The
$V_w$ velocities in Table~1 are the pre-TS velocities.  The
ratio of the two is a measure of the compression ratio ($\eta$) of
the TS.  Mathematically, $\eta=V_w/[V_{rad}-V({\rm Mg II})]$.
The $\eta$ values of the three detected astrospheres are listed
in Table~2.

     Assuming the absorption line widths are dominated by
thermal broadening, the Doppler parameters (in km~s$^{-1}$) will be
related to temperature by $b^2=0.0165T/A$, where $A$ is the atomic
weight of the species in question.  The last column of Table~2 lists
temperatures computed from this equation using the Mg~II Doppler
parameters (with $A=24$).

     If the TS is radiative, some of the astrospheric absorption
may be coming from the intermediate radiative relaxation zone of the
shock, rather than being entirely from post-TS material that has
reached its final flow velocity and temperature.  The effects of
velocity gradients in this region would be to broaden the
absorption line and shift the line centroid blueward of the true
terminal post-TS velocity.  This would mean that $\eta$ in Table~2
could actually be an underestimate and T an
overestimate.  However, velocity gradients in the relaxation region
should also induce asymmetries in the absorption profiles, for which
we do not see any evidence, which is an argument in favor of the
absorption being dominated by post-relaxation zone material.

     At the TS, the kinetic energy of the stellar wind is
partly converted to thermal energy in the post-TS material.  Thus,
there should be a correlation between $V_w$ and the temperatures
listed in Table~2.  We can be more specific using the conservation of
momentum equation of the Rankine-Hugoniot shock jump relations, which
says
\begin{equation}
P_1+\rho_1 v_1^2 = P_2+\rho_2 v_2^2,
\end{equation}
where $P_1$, $\rho_1$, and $v_1$ are the pressure, density, and
velocity upstream of the shock; and $P_2$, $\rho_2$, and $v_2$ are the
corresponding quantities downstream.  An MHD version of equation (3)
would include $B^2/8\pi$ terms, but such terms should not be
significant here.  In the strong shock limit
appropriate here, $P_1$ is negligible and
$\eta=\rho_2/\rho_1=v_1/v_2=4$ for a simple hydrodynamic
shock \citep[e.g.,][]{fhs92}.  Thus, equation (3) reduces to
\begin{equation}
P_2=\frac{3}{4} \rho_1 v_1^2.
\end{equation}
The post-TS pressure and temperature are related by $P_2=n_2k_BT_2$,
where $k_B$ is the Boltzmann constant and $n_2$ the number
density.  For a pure hydrogen fluid,
$\rho_1=m_p n_1$, where $m_p$ is the proton mass.  Plugging these
relations into equation (4) and noting that $n_2/n_1=\eta=4$ leads to
\begin{equation}
T_2=22.7 v_1^2,
\end{equation}
if $v_1$ is in km~s$^{-1}$ units.

     In Figure~5, we plot this expected relation between stellar
wind speed and post-TS temperature, which is compared with the actual
measurements for the three detected astrospheres.  We do see an
increase in temperature with $V_w$, but the measured temperatures are
lower than predicted.  The explanation for the discrepancy lies in the
assumption of a simple hydrodynamic shock in deriving equation (5),
which requires $\eta=4$ in the strong shock limit.  However, from
Table~2 we already know $\eta>4$ for all three detected astrospheres.
This is evidence for a dissipative shock, as suggested before for
$\alpha$~Tau, where the astrospheric absorption could only be
reproduced when the TS was modeled as a radiative shock \citep{bew07}.
The dashed line is representative of the initial post-TS
temperature, but radiative cooling decreases this temperature,
resulting in additional compression.  It is worth noting that the star
with the lowest $\eta$, $\gamma$~Eri, is not surprisingly closest to
the expected temperature in Figure~5.  For $\gamma$~Eri, less
radiative cooling and associated compression is necessary to reach the
observed temperature.  This is explored further with a full hydrodynamic
model of the $\gamma$~Eri astrosphere in the next section.
\begin{figure}[t]
\plotfiddle{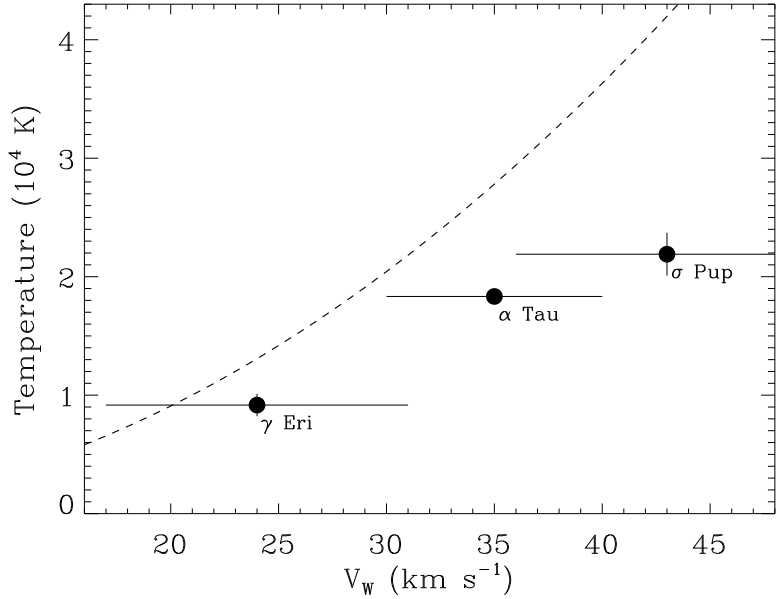}{3.0in}{90}{50}{50}{180}{-40}
\caption{The post-TS temperature estimated from the astrospheric
  Mg~II lines in Table~2 is plotted versus the stellar wind velocity
  estimates from Table~1.  The dashed line is the relation expected
  for a simple hydrodynamic shock at the strong shock limit, assuming
  a pure hydrogen fluid.}
\end{figure}

     A final question to address in this section concerns the issue
of why only three of the red giant targets in Table~1 show the
astrospheric absorption signature.  The $\theta$ values of the
three archival nondetections ($\alpha$~Boo, $\gamma$~Dra, and
$\gamma$~Cru) demonstrate that we are not looking through the astrotail,
and therefore we may not have sufficient Mg~II column densities
along the lines of sight to these stars to detect astrospheric
absorption.  This, however, is not true for the 7 new targets that
did not yield detections.

     One problem for the three M3-M5 giants
in that list ($\mu$~Gem, $\tau^4$~Eri, and HD~120323) is that there is
very little flux near the stellar rest frame to provide a suitable
background for the astrospheric signature, making it harder to
detect.  This is presumably due to a combination of low wind
speed and high mass loss rate.  As discussed above with regards
to $\gamma$~Eri, the low wind speed also reduces the heating at
the TS, thereby reducing radiative cooling and post-TS compression,
which also reduces astrospheric detectability.

     Three of the K2-K4 giants among the nondetections
(HD~66141, $\alpha$~Tuc, and HD~87837) have
winds weak enough that there is no deep, broad wind absorption
feature in the Mg~II lines.  Perhaps for these stars the winds
are not strong enough to carve out astrospheres that are sufficiently
large to yield enough post-TS Mg~II column density
for an astrospheric absorption detection.  This does not explain
the $\theta$~CMa nondetection, though.  The $\theta$~CMa nondetection
is the hardest to explain, as there is strong wind absorption in
the Mg~II profiles, there is plenty of flux near the stellar rest
frame to provide background for the astrospheric signature, and
the $V_w$ and $\theta$ values for the star are both large enough
to expect a detection.  In any case, given that only 2 of the 9
stars in our new survey yielded astrospheric detections, despite
an effort to choose targets likely to provide detections, it is
clear that detectable astrospheric absorption signatures are
the exception and not the rule for red giants, even in high
quality spectra.

\section{The $\gamma$~Eri Astrosphere}

     We compute a hydrodynamic model of the
$\gamma$~Eri astrosphere in much the same manner as we modeled the
$\alpha$~Tau astrosphere \citep{bew07}.  This is a two-fluid
model of a type used to model the heliosphere, in which the plasma
and neutrals are treated as two separate, distinct fluids that are
allowed to interact through charge exchange processes
\citep{hlp95}.  A first step in the modeling process is to
define the boundary conditions for the model, specifically the
stellar wind boundary conditions at 1~AU from the star, and the
parameters for the surrounding ISM.

     For the stellar wind, we model the Mg~II wind absorption feature
using techniques used in the past \citep{gmh95,bew07},
leading to a mass loss rate
estimate of $\dot{M} \simeq 3\times 10^{-11}$ M$_{\odot}$~yr$^{-1}$
and a terminal velocity of $V_w=20$ km~s$^{-1}$.  These values
can be used to infer densities at 1~AU from the star, representing
the inner boundary condition of the hydrodynamic model.  The model
only considers hydrogen, expected to be the dominant constituent by
number and mass.  The ionization state of hydrogen is very uncertain.
It is expected to be mostly neutral at the $T\sim 10^4$~K temperature
of the wind, but some degree of ionization may be present.  We
assume a 10\% ionization.  We assume a temperature of 7500~K for
the wind at 1~AU, though this is actually an unimportant parameter
given that the flow is very supersonic.

     Turning to the interstellar boundary conditions, the ISM flow
velocity and direction represented by the $V_{ISM}$ and $\theta$
values from Table~1 are necessary constraints.  As for density and
temperature appropriate for hot LB plasma, we assume a temperature and
proton density of $T_{\infty}=5\times 10^{5}$~K and
$n_{\infty}({\rm H^{+}})= 0.01$ cm$^{-3}$, respectively,
as assumed for Model~8 of the $\alpha$~Tau astrosphere from \citet{bew07}.
There are small, partly neutral clouds within the LB,
including the LIC surrounding the Sun, but we assume $\gamma$~Eri will
be surrounded by the more typical hot LB plasma.

\begin{figure}[p]
\plotfiddle{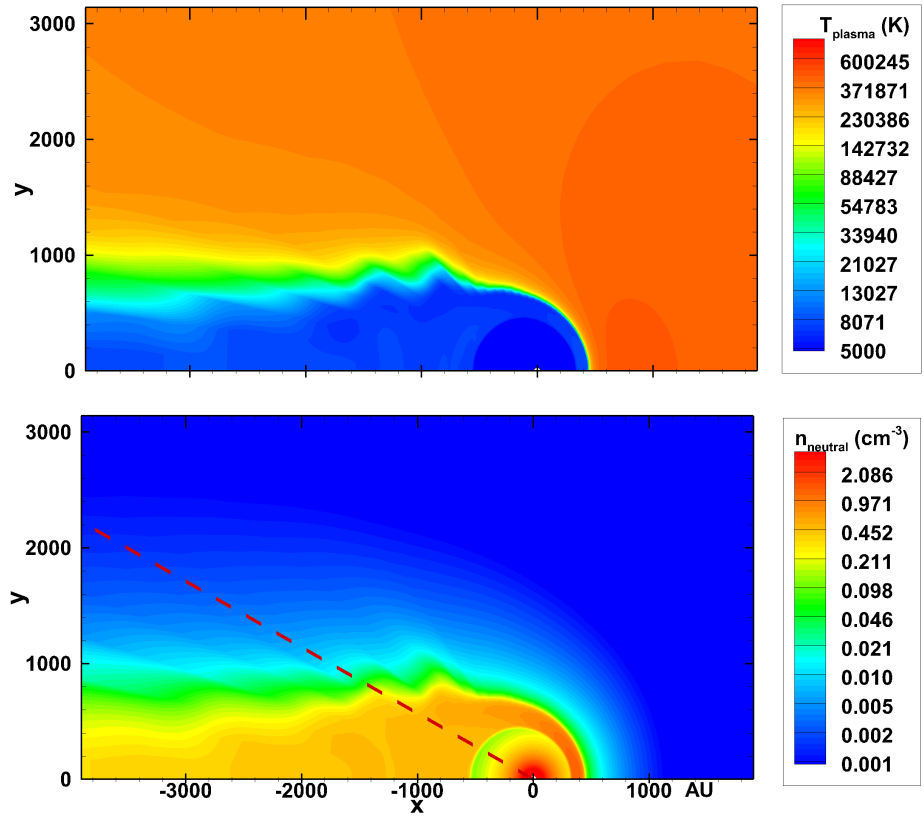}{4.0in}{0}{55}{55}{-240}{-20}
\caption{Hydrodynamic model of the $\gamma$~Eri astrosphere, with
  top panel showing temperature and the bottom panel showing number
  density.  The dashed line in the bottom panel indicates our line of
  sight to the star.}
\end{figure}
\begin{figure}[h]
\plotfiddle{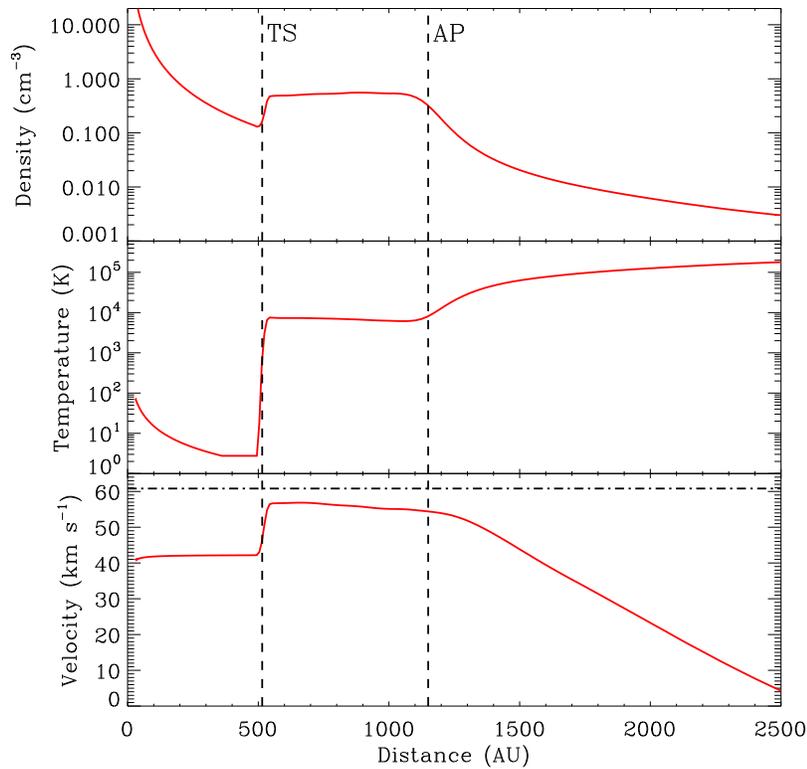}{3.6in}{90}{60}{60}{240}{-40}
\caption{Neutral hydrogen density, temperature, and velocity along the
  line of sight to the star shown in Figure~6.  The dot-dashed line in the
  bottom panel is the rest frame of the star.  The positions of the
  termination shock (TS) and astropause (AP) are noted.}
\end{figure}
     Figure~6 shows the resulting hydrodynamic model of $\gamma$~Eri.
The figure shows maps of temperature and density.  The stellar wind
expands radially from the star until it hits the roughly circular
TS, at a distance of about 350~AU in the upwind direction and
about 550~AU in the downwind direction.  The TS is seen most clearly
in the density plot in Figure~6.  The heated, compressed, and
decelerated post-TS wind material is partly deflected downwind towards
the astrotail, separated from the plasma flow of the ISM by a
parabola-shaped ``astropause'' (analogous to the heliopause),
although the large mean free path of the neutrals allows them to
cross this boundary and create the halo of neutrals around it,
in green in the density plot of Figure~6.  Unlike most heliospheric
models, the ISM flow does not pass through a bow shock, as the ISM
flow is not supersonic due to the high temperature of the assumed
LB plasma.

     Figure~7 shows the traces of H~I density, temperature, and
velocity along our $\theta=146^{\circ}$ line of sight to $\gamma$~Eri.
The locations of the TS and astropause in this direction are noted.
It is the post-TS material in between the TS and astropause that
will be responsible for the astrospheric Mg~II absorption feature,
as that is where the density is highest.  Computing Mg~II absorption
from the hydrogen parameters in Figure~7 requires that we assume that
Mg~II is the dominant ionization state of Mg, and we assume a
solar Mg abundance \citep{ng98}.

     The Mg~II absorption predicted by the model is compared with
the data in Figure~8a.  The model significantly underpredicts the
amount of absorption, to the extent that it is difficult to see
the difference between the predicted absorption and the assumed
background flux above the absorption.  Thus, Figure~8b shows a
different version of this plot, showing the opacity across the line
profile.  The model is quite successful in predicting the location and
width of the absorption, and if the line opacity is arbitrarily
increased by a factor of 8, the predicted absorption actually fits the
data reasonably well.  The model in Figure~6 is actually more
successful in reproducing the observed astrospheric absorption than
the hydrodynamic $\alpha$~Tau models, most of which underpredicted the
amount of absorption even more severely, and all of which placed the
absorption much too far from the stellar rest frame.  Treating the
$\alpha$~Tau TS as a radiative shock was necessary to increase the
amount of absorption and to shift it closer to the stellar rest frame,
although even with the radiative shock treatment there was still
difficulty predicting a sufficient amount of absorption
\citep{bew07}.

     For $\gamma$~Eri, the predicted absorption is slightly
blueshifted from its observed location, but not by much.  This
is consistent with the $\eta=5.9\pm 1.7$ value reported in
Table~2, which indicates a compression ratio slightly higher
that the $\eta=4$ value of a strong, hydrodynamic shock, but
not by very much, unlike $\alpha$~Tau.  The need to include
radiative cooling is therefore much reduced for $\gamma$~Eri.
Including radiative cooling in the $\gamma$~Eri model,
which we will not try here, would presumably improve
agreement with the data somewhat, but it is unlikely to be
the main reason the model is greatly underestimating
the amount of absorption, considering that $\eta$ is not too
much higher than 4 for $\gamma$~Eri.
\begin{figure}[t]
\plotfiddle{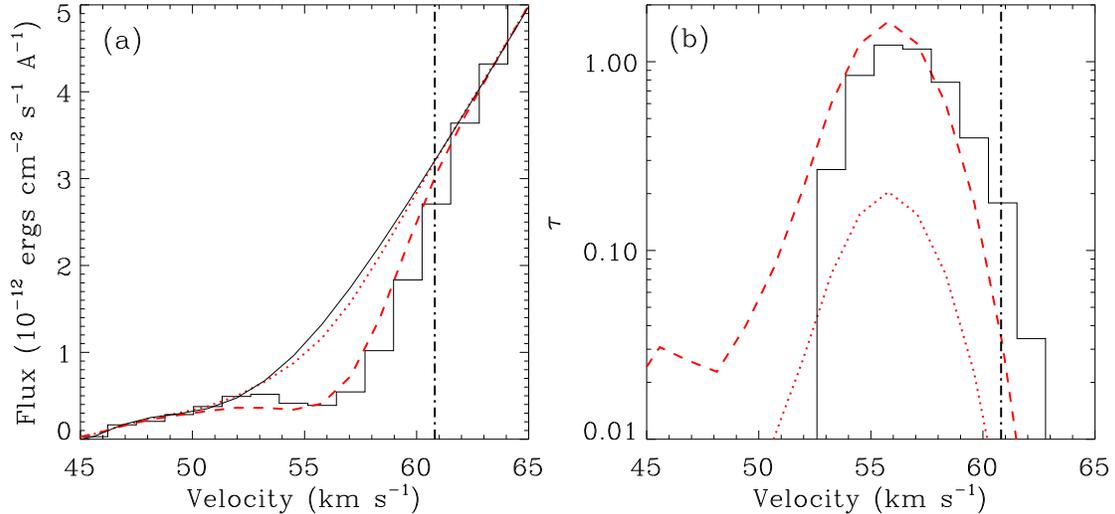}{2.5in}{90}{65}{65}{250}{-95}
\caption{(a) Comparison of observed astrospheric Mg~II absorption
  towards $\gamma$~Eri (histogram) with the absorption predicted by
  the model in Figure~6 (dotted line).  The dashed line is the
  predicted absorption after opacities are arbitrarily increased by a
  factor of 8 to better fit the data.  (b) Comparison of observed
  astrospheric Mg~II line opacity with that predicted by the Figure~6
  model (dotted line), with the dashed line being the predicted
  opacity increased by a factor of 8 to better fit the data.}
\end{figure}

     Two other possible reasons for the absorption underestimate
that are worth considering are: 1. Our assumed Mg abundance in
the stellar wind is too low, and 2. The ISM pressure that
we are assuming may be too low.
As for the first explanation, we are not aware of any precise
measurement of a photospheric Mg abundance for $\gamma$~Eri,
but even if such a measurement did exist, the Mg abundance in
the stellar wind could be different.  Even in the solar wind,
the abundances of elements with low first ionization potential
can be higher than photospheric abundances by a factor of 4
\citep{rvs95}.  Thus, it is plausible that simply
assuming a solar photospheric Mg abundance as we have could be a
significant underestimate.

     As for the second suggestion, the effects of a high ISM
pressure on astrospheric absorption were explored at length
in the analysis of the $\alpha$~Tau astrosphere \citep{bew07}.
Increasing the ISM density pushes the TS closer to the star,
where wind densities are higher, leading to the post-TS densities
being higher, thereby producing more absorption.  In contrast, the
amount of astrospheric absorption was unaffected by increases in
stellar wind density, because the wind density increase was offset by
the effect of pushing the TS farther from the star, where densities
are lower.  The astrospheric absorption features are therefore
potentially useful as ISM pressure diagnostics.  The ISM
pressure that we are assuming in the $\gamma$~Eri model ($P/k_B=5000$
cm$^{-3}$~K) is consistent with measurements of pressures of neutral
clouds within the LB \citep{ebj02}, but it may be a little low based
on recent assessments of hot plasma within the LB, which suggest
$P/k_B=10,700$ cm$^{-3}$~K \citep{sls14}.  Thus, it is possible
that an underestimate of the ISM pressure could be part of the
explanation for the underestimate of astrospheric absorption.

\section{The $\sigma$~Pup Stellar Wind and Astrosphere}

     Interpretations of $\sigma$~Pup's Mg~II lines and attempts to
model its wind and astrosphere are complicated by its binarity.  As
noted in section~2, the K5~III giant has a G5~V companion, with an
orbital period of 257.8~days.  The half-amplitude of the orbital
motion is $K1=18.6$ km~s$^{-1}$ \citep{dp04}.  This is
comparable to the wind speed, so the orbital motion will impart a
significant increase in this speed in the direction of orbital motion,
and a corresponding decrease in speed in the opposite direction.
During the course of one 257.8~day orbit the $V_w=43$ km~s$^{-1}$
stellar wind would travel about 6.5~AU, or about 32 stellar radii.
The region around the binary on these distance scales will involve a
complex merger of high speed and slow speed streams, with uncertain
consequences for the effective final terminal wind speed, although it
is expected that the high and slow speed streams will have merged into
a more homogeneous outflow before the TS is reached at a distance of
$\sim 1000$~AU.  Thus, the systemic center-of-mass velocity quoted as
$V_{rad}$ in Table~1 ($V_{rad}=87.3$ km~s$^{-1}$) is definitely a better
reference velocity to use when computing the post-TS velocity and
$\eta$ than the actual stellar radial velocity at the time of the
observation, which inspection of narrow H$_{2}$ lines in the E140M
spectrum suggests is $V_{rad}=103$ km~s$^{-1}$.  The observed
astrospheric absorption is certainly close to the former and not the
latter.

     More questionable is our use of $V_{rad}=87.3$ km~s$^{-1}$
as the reference velocity for computing $V_w$.  Which $V_{rad}$ value
is appropriate for estimating $V_w$ depends on whether the wind
absorption feature is entirely from the part of the wind close
to the star and moving with the star, or whether there is
absorption from material farther from the star emitted when the
star had a different radial velocity.  By assuming
$V_{rad}=87.3$ km~s$^{-1}$ we are implicitly assuming the latter,
but the former is quite possible.  If $V_{rad}=103$ km~s$^{-1}$
is more appropriate, our $V_w$ estimate would increase
to $V_w=59\pm 7$, which would make $\sigma$~Pup look discrepant
in Figure~2c, albeit still not as discrepant as $\gamma$~Dra.
If $\sigma$~Pup's Mg~II lines could be monitored over the course
of an orbit, we could see if the stellar wind absorption precisely
follows the star or not, thereby resolving this particular issue.
Such monitoring has been done in the optical for the Ca~II lines of
the spectroscopic binary $\mu$~UMa (M0~III+?), with a 230 day period,
where it is found that the wind absorption does not seem to follow the
radial velocity shifts of the red giant primary \citep{dr77}.  This
supports our use of the systemic $V_{rad}=87.3$ km~s$^{-1}$ value as
the reference velocity in computing $V_w$, instead of
$V_{rad}=103$ km~s$^{-1}$.

     A final complicating factor that should be mentioned concerns
the presence of the G5~V companion star, which will
be embedded within the wind emitted by the red giant.  Gravitational
effects on the relatively slow red giant wind could possibly be
significant, as could interactions between the winds of the two stars.
If the companion star's wind is like that of the Sun, it may be too
weak compared with the massive red giant wind to have any effect on
the red giant wind, but this is far from certain.
In any case, with all these complications in mind, we choose not
to compute a $\sigma$~Pup astrospheric model at this time.

\section{Summary}

     We have analyzed the chromospheric Mg~II h \& k lines of
K2-M5~III stars observed by HST, consisting of 9 observations
obtained as part of a new HST red giant survey, and 4
archival targets.  Our findings are summarized as follows:
\begin{enumerate}
\item The Mg~II line profiles of all 13 stars in our sample show
  evidence for stellar wind absorption, but for three of the
  K2-K4~III stars the effect is only an induced asymmetry in
  the line profile rather than a deep wind absorption feature.
\item Measured Mg~II surface fluxes are very tightly correlated
  with spectral type and photospheric temperature, consistent with the
  idea that K2-M5~III stars redward of the coronal dividing line are
  all emitting at a basal flux level.  The Mg~II k/h flux ratio
  increases towards later spectral types.
\item Wind speeds estimated empirically from the Mg~II spectra
  correlate with spectral type and photospheric temperature, with $V_w$
  decreasing from $V_w\approx 40$ km~s$^{-1}$ at K2~III to
  $V_w\approx 20$ km~s$^{-1}$ at M5~III.
\item There are 2 new detections of astrospheric absorption among
  the recently observed stars ($\gamma$~Eri and $\sigma$~Pup), for a
  total of 3 in our sample, including the previous detection towards
  $\alpha$~Tau.  However, the limited number of new detections
  indicates that detectable astrospheric absorption in UV lines is not
  a common phenomenon.  For both $\gamma$~Eri and $\sigma$~Pup we
  detect astrospheric Fe~II $\lambda$2600.2 absorption in addition
  to the Mg~II signature, and for $\sigma$~Pup, astrospheric absorption is
  also observed in C~II $\lambda$1334.5.  These are the
  first detections of red giant astrospheres in lines other than Mg~II.
\item Analysis of the astrospheric absorption leads to
  measurements of TS compression ratio and post-TS temperature
  for the three detected astrospheres.  The temperatures
  are correlated with $V_w$.  However, the $T=(0.9-2.2)\times 10^4$~K
  post-TS temperatures are too low and the $\eta=6-18$ compression
  ratios too high according to the Rankine-Hugoniot shock jump
  conditions, providing further evidence that red giant termination
  shocks are radiative shocks rather than simple hydrodynamic shocks.
\item We compute a hydrodynamic model of the $\gamma$~Eri
  astrosphere, which is the one with the TS compression ratio
  ($\eta=5.9\pm 1.7$) closest to the strong hydrodynamic shock value
  of $\eta=4$.  Not surprisingly, this model has less difficulty
  reproducing the observed absorption than was the case for a past
  study of the $\alpha$~Tau astrosphere \citep{bew07}, for which
  $\eta=17.5\pm 3.1$.  The $\gamma$~Eri model places the absorption at
  about the right velocity and with about the correct width, but it
  underpredicts the Mg~II opacity by a factor of 8.  This might be due
  to either an underestimate of the Mg abundance in the stellar wind,
  or an underestimate of the ISM pressure surrounding the star.
\end{enumerate}

\acknowledgments

We would like to thank Drs.\ Seth Redfield and Alexander Brown
for contributing to the acquisition and analysis of the Ca~II
spectra.  Support for HST program GO-13462 was provided by NASA
through an award from the Space Telescope Science Institute,
which is operated by the Association of Universities for
Research in Astronomy, Inc., under NASA contract NAS 5-26555.
This research has made use of the SIMBAD database, operated at
CDS, Strasbourg, France.

\end{document}